\documentclass[12pt,preprint]{aastex}

\newcommand{\ts}{\thinspace}

\shorttitle{COSMOS-Pairs}
\shortauthors{Kartaltepe et al.}

\begin{document}

\title{Evolution of the Frequency of Luminous ($\geq L_{\rm V}^\star$) Close Galaxy Pairs at $z < 1.2$ in the COSMOS Field}

\author{J. S. Kartaltepe\altaffilmark{1}, D. B. Sanders\altaffilmark{1}, N. Z. Scoville\altaffilmark{2}, D. Calzetti\altaffilmark{3}, P. Capak\altaffilmark{2}, A. Koekemoer\altaffilmark{3}, B. Mobasher\altaffilmark{3}, T. Murayama\altaffilmark{4}, M. Salvato\altaffilmark{2}, S. S. Sasaki\altaffilmark{4,5}, and Y. Taniguchi\altaffilmark{4,5}}

\altaffiltext{$\star$}{Based on observations with the NASA/ESA {\em Hubble Space Telescope}, obtained at the Space Telescope Science Institute, which is operated by AURA Inc, under NASA contract NAS 5-26555; also based on data collected at : the Subaru Telescope, which is operated by the National Astronomical Observatory of Japan; Kitt Peak National Observatory, Cerro Tololo 
Inter-American Observatory, and the National Optical Astronomy Observatory, which are operated by the Association of Universities for Research in Astronomy, Inc.(AURA) under cooperative agreement with the National Science Foundation; and the Canada-France-Hawaii Telescope operated by the National Research Council of Canada, the Centre National de la Recherche Scientifique de France and the University of Hawaii.}  

\altaffiltext{1}{Institute for Astronomy, 2680 Woodlawn Dr., University of Hawaii, Honolulu, Hawaii, 96822, email: jeyhan, sanders@ifa.hawaii.edu}

\altaffiltext{2}{California Institute of Technology, MC 105-24, 1200 East California Boulevard, Pasadena, CA 91125}

\altaffiltext{3}{Space Telescope Science Institute, 3700 San Martin Drive, Baltimore, MD 21218}

\altaffiltext{4}{Astronomical Institute, Graduate School of Science, Tohoku University, Aramaki, Aoba, Sendai 980-8578, Japan}

\altaffiltext{5}{Physics Department, Graduate School of Science, Ehime University, 2-5 Bunkyou, Matuyama, 790-8577, Japan}

\begin{abstract}
We measure the fraction of luminous galaxies in pairs at projected separations of $5-20${\ts}kpc out to $z=1.2$ in the COSMOS field using ACS images and photometric redshifts derived from an extensive multiwavelength dataset. Analysis of a complete sample of 106,188 galaxies more luminous than $M_V =-19.8$ ($\sim L_V^\star$) in the redshift range $0.1 < z < 1.2$ yields 1,749 galaxy pairs.   These data are supplemented by a local ($z=0-0.1$) value for the galaxy pair fraction derived from the {\it Sloan Digital Sky Survey} (SDSS).  After statistically correcting the COSMOS pair sample for chance line-of-sight superpositions, the evolution in the pair fraction is fit by a power law $\propto (1+z)^{n=3.1\pm0.1}$. If this strongly evolving pair fraction continues out to higher redshift, $\sim 50\%$ of all luminous galaxies at $z\sim2$ are in close pairs. This clearly signifies that galaxy mergers are a very significant and possibly dominant mechanism for galaxy evolution during the epoch of galaxy formation at $z=1$ to 3. 

\end{abstract}


\section{INTRODUCTION}

Galaxy mergers play an important role in galaxy formation and evolution. Not only do they change the number density of galaxies in the universe and provide a method for mass accretion, but they also significantly alter the shapes and morphologies of galaxies, induce bursts of star formation, and provide material for accretion onto black holes. Current models of galaxy formation suggest that galaxy mergers must have been much more common in the past than they are now \citep{too77,bar92}. 

Though the significant role of merging has long been recognized, quantitative estimates of merging rates have varied greatly. Derived merger rates depend strongly on the redshift range surveyed, the wavelength band used for imaging, and the limiting magnitude of the survey. In addition, all previous studies at $z \geq 0.1$ have involved relatively small galaxy samples - typically a few tens up to a few hundred merger candidates. Most of the early studies use pair fraction as a surrogate for the merger rate \citep[e.g.,][]{zep89,bur94,woo95,yee95,neu97,pat02,wu98,car00,lef00,lin04}; however, the pair fraction may not translate so simply \citep[e.g.,][]{abr99}. Instead, it samples galaxy pairs at different stages of interaction and assumes that that they will all eventually merge. Although most close galaxy pairs may ultimately merge, derivation of the actual merger rate from the pair fraction requires an understanding of the merging timescale which is likely to be strongly dependent on both the observed separation and the assumed orbital eccentricity.

Several recent studies use morphological criteria to select disturbed systems in the advanced stages of merging \citep[e.g.,][]{lef00,con03,lav04,lot06}. While this method is more likely to yield objects which have just recently merged or are very likely to merge, it has its own deficiencies: information on the progenitor galaxies is often lost and the morphological selection can be very redshift-dependent due to surface brightness dimming and morphological K-corrections. The morphological appearance of galaxies can be quite dependent on the observed wavelength, particularly as one observes more and more of the rest frame UV wavelengths. The galaxy morphologies then become increasingly dominated by irregular areas of recent star formation, rather than galactic mass structures. When the star formation is particularly intense and clumped, the blue-UV morphology may then mimic the disturbed appearance of interacting systems. 

Given the differences and problems listed above, merhaps it is not surprising then that both techniques -- pair fraction and morphology -- have yielded a very broad range of power-laws $((1+z)^m)$ for the merger rate -– from $m = 0$ (no evolution) to $m \sim 5$ (strong evolution) as reported in the literature (see \S 5.2).

The Cosmic Evolution Survey \citep[COSMOS:][]{scoville_cosmos_overview} offers a new and unique opportunity to study the galaxy merger rate versus  redshift.  The availability of HST imaging and galaxy catalogs for the full 2 square degree COSMOS field  (Scoville et al. 2007c, Koekemoer et al. 2007, Leauthaud et al. 2007) provides an enormous sample of galaxies, while extensive ground-based follow-up imaging of the full field in the optical and near-infrared provides accurate photometric redshifts \citep{mobasher_cosmos_photz}. Here we make use of these data to investigate the evolution of the galaxy pair fraction, taking advantage of the high sensitivity and redshift accuracy of the COSMOS catalogs.  Specifically we use a sample of 106,188 $\geq L^\star$ galaxies to derive the fraction of close galaxy pairs (at $5-20${\ts}kpc separation) out to $z = 1.2$.   

Most earlier studies have used spectroscopic redshifts to confirm that galaxies close in position on the sky are indeed at similar redshifts and therefore possibly gravitationally bound. These studies have  typically probed projected separations less than a few tens of kiloparsecs with $\Delta V < 1000${\ts}km/s.  However, the requirement for spectroscopic redshift confirmation usually results in very limited sample sizes, and since the full 3-d velocity is still not known even for the spectroscopic samples, one still doesn't know that the pair is indeed bound. Thus the major utility of the spectroscopic redshifts is to greatly reduce line-of-sight contamination or chance superposition. However, since the line of sight contamination can be statistically estimated, corrected for, and kept at a low level (by avoiding very large impact parameters), we make use of photometric redshifts; these have lower accuracy than spectroscopic redshifts but yield vastly larger samples. The photometric redshift sample has the added benefit that it is also free of the fiber collision problems and incompleteness which often plague spectroscopic samples.

The structure of this paper is as follows. The data sets used for this project are described in \S 2 and the selection criteria used to detect the pairs are discussed in \S 3. Our results are presented in \S 4 and discussed in \S 5 where we compare results with those of previous studies on the pair fraction and merger rate evolution. Our conclusions are presented in \S 6. The cosmological parameters used throughout this paper are $h=0.7$, $\Omega_{\Lambda}=0.7$, and $\Omega_{m}=0.3$.

\section{THE DATA}
The imaging data used for this study were obtained as part of the HST-COSMOS project \citep{scoville_cosmos_hst}. COSMOS originated as an {\it HST} Treasury program imaging an $\sim$2{\ts}\sq$^\circ$\  equatorial field with the Advanced Camera for Surveys (ACS), using the F814W filter (I-band). This is the largest contiguous field ever observed by HST. The field size was chosen to sample the broad range of large scale structure (LSS) and thereby reduce the problem of cosmic variance that plagues smaller survey fields \citep{scoville_cosmos_overview}. Details of the ACS images, including their calibration and reduction, are given in Koekemoer et al. (2007). In addition to the ACS images, we used the ACS source catalog that is complete to $I < 26$ \citep{lea_cosmos_catalog}.

The COSMOS project has also obtained extensive multiwavelength ground and space-based observations covering the entire field. A complete description of the COSMOS data sets can be found in \citet{scoville_cosmos_overview}. A ground-based photometry catalog was produced using imaging data taken with Subaru-SuprimeCam ($B V g^\prime R i^\prime z^\prime$ NB816), CFHT-Megacam ($u^*$,$i^*$), CTIO-ISPI ($K_{\rm s}$ band), NOAO Kitt Peak-4m ($K_{\rm s}$ band) and SDSS ($u^\prime g^\prime r^\prime i^\prime z^\prime$). Details of the observations, data reduction and catalogs can be found in Aussel et al. (2007), \citet{capak_cosmos_catalog}, and \citet{taniguchi_cosmos_subaru}.  Using the ground-based photometry, \citet{mobasher_cosmos_photz} calculated photometric redshifts for approximately one million galaxies within the COSMOS field. These photometric redshifts have an accuracy of $dz/(1+z) \sim 0.03$ out to a redshift of $z=1.2$ \citep{mobasher_cosmos_photz}.  The photometric redshift catalog is crucial for our study, enabling a sorting of the galaxies in redshift 
bins to reduce contamination by chance, line-of-sight pairs and hence to yield a reliable sample of  galaxy pairs.

At low redshift ($z < 0.2$), the COSMOS field encompasses too small a volume to provide a meaningful esitimate of the local galaxy pair fraction. Therefore, to estimate the ``local" ($z = 0-0.1$) fraction of galaxy pairs, we make use of the published catalog of Sloan Digital Sky Survey (SDSS) galaxy pairs by \citet{all04}. This catalog samples a volume in the local universe nearly as large as that in the COSMOS field at $z=1$. 

\section{ANALYSIS}

\subsection{Galaxy Sample Selection Criteria}
The primary criterion for galaxy selection was an absolute luminosity cutoff corresponding to a $\sim${\ts}$L_V^\star$ galaxy, specifically $M_V=-19.8$ \citep{bla01}.  This was done partly for practical reasons in that galaxies above this luminosity have apparent I band magnitudes of $< 26$ and are still easily visible out to the redshift limit of this study, $z=1.2$, thus eliminating the need to correct for incompleteness at higher redshifts. More importantly, it is likely that pairs of galaxies brighter than $L_V^\star$ will become components of approximately equal mass mergers since most of these galaxies are near the ``knee" of the luminosity function (since above the ``knee", the 
number density of galaxies drops steeply). Lastly, we note that mergers between galaxies with luminosities $\geq L^*$ are likely to be responsible for some of the most luminous and rapid phases in galaxy evolution, including ultraluminous infrared galaxies \citep[ULIRGs: e.g., ][]{kim98,das06} and quasi-stellar objects \citep[QSOs: e.g.,][]{vei06,guy06}. To ensure that all the galaxies in our sample have reliable photometric redshifts we required that each be detected in at least four bands, including $K_{\rm s}$ band to guarantee a wide wavelength coverage. Applying the above selection criteria to the 438,356 entries in the ground-based catalog yielded a sample of 106,188 galaxies $\geq L_V^\star$ across the entire COSMOS field. Selecting all galaxies with $z<1.2$ yields a final sample of 59,221 galaxies used to search for pairs.

For studies which examine the pair fraction as a function of redshift it is critical to impose a lower limit to the galaxies' luminosities (or even better their masses) rather than an apparent magnitude cutoff.  If the various redshift slices probe to different limiting masses (or luminosity) on the galaxy mass function, then the slice going to lower mass will have more objects per unit volume. This  will obviously result in a larger pair fraction, i.e. a larger number of companions within a specified radius, independent of any evolution \citep[cf.  ][]{ber06}.  For this reason we attempt to encompass the same ``classes" of objects across the full redshift range of our study and do this by adopting a fixed cutoff in $L_V^\star$. We point out in \S 5.3 that the use of a fixed cut in luminosity is more appropriate than a passive evolution cut (i.e. lower luminosity at low redshift). The vast majority of the merging systems are star-forming disk galaxies with ongoing star formation even to the lowest redshifts in our sample (\S 5.1); they are not passively evolving.

\subsection{Identification of Galaxy Pairs}

To identify possible galaxy pairs, we searched for the {\bf nearest-neighbor} galaxy for each galaxy in the sample -- using photometric redshifts to limit the difference in redshift to $\Delta z \leq \pm0.05$ ($\simeq$ the typical uncertainty in the photometric redshift estimates). Thus both the primary and its possible companion must have $L \geq L_V^\star$  and the most probable redshifts, from the photometric redshift fitting, must be within 5\%. From computed probability distributions derived from the photometric redshift fitting, we estimate that $\sim$90\% of pairs would be found. This is likely to change slightly (ranging from $\sim$80-95\%) with redshift since the error in the photometric redshift estimate depends on redshift. However, as discussed in \S 3.3, we calculate the expected random distribution using precisely the same numbers of galaxies in each redshift slice and therefore this small variation is corrected for. We also limited the range of projected separations to minimize contamination from chance, line-of-sight pairs.  

Galaxy pairs were identified in two ways -- a primary search yielding over 95\% of the sample was done using the ground-based photometric redshift catalog \citep{mobasher_cosmos_photz}; this was supplemented with a search for very close pairs using the ACS-based catalog \citep{lea_cosmos_catalog}. The ground-based catalog was the primary catalog used for our analysis 
since it contains the majority of the sample and immediately provides redshifts and absolute magnitudes for all objects. However, the resolution of the ground-based images mean that a small fraction of the most closely separated galaxy pairs are missed in the photometric redshift-based search.  To ensure that we have indentified all pairs at these close separations, we then used the catalog extracted from the ACS images to search for close companions which are blended and appear as a single entry in the photometric redshift catalog. (For these close objects, we adopted 
the single redshift from the photometric redshift catalog.) Any ``ACS-only" pairs found in the secondary search which were already picked up in the ground based catalog were of course excluded. ``ACS-only" pairs for which the photometric redshift fit was poor were also dropped from further consideration since these could be superpositions at different redshifts. Each of the ``ACS-only" galaxy pairs was also visually inspected to remove ``artificial pairs" which might have been present in the ACS catalog -- for example, features of the same nearby galaxy, such as a bright spiral arm or residual cosmic rays.   

\subsection{Correction for Line-of-Sight Superpositions}

\begin{figure}
\plotone{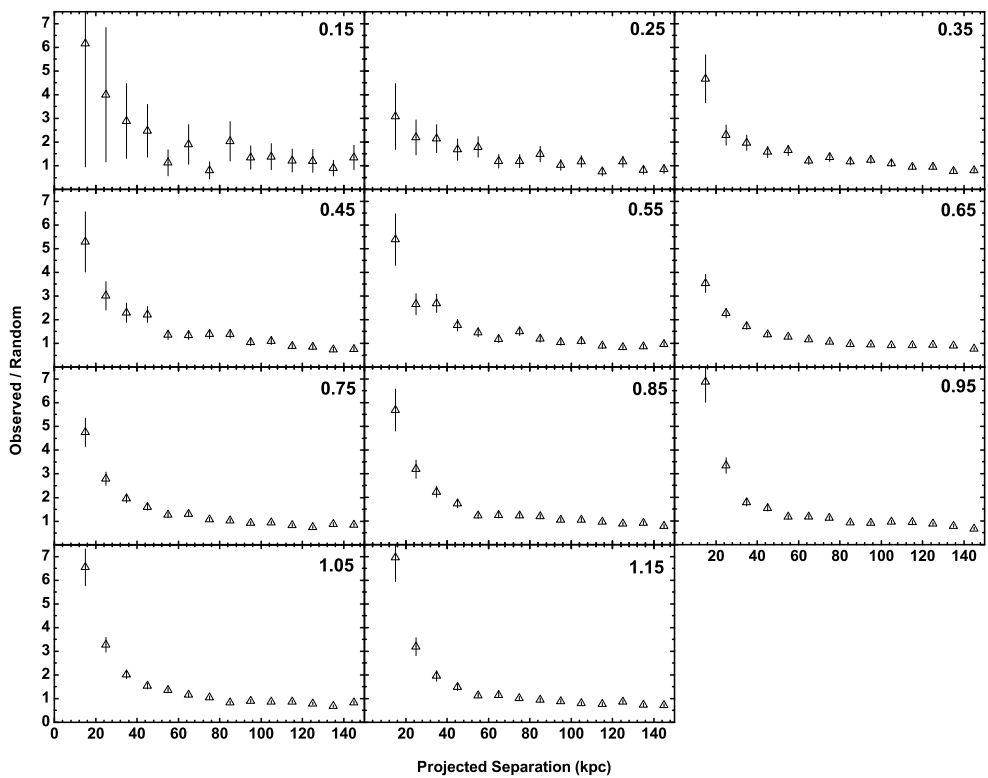}
\vskip 0.4truein
\caption{Ratio of observed to random pairs versus pair projected separation in $\Delta z=0.1$ redshift bins between $z=0.1$ and $1.2$ measured for galaxies from the ground-based photometric redshift catalog. This distribution represents the likelihood that a galaxy has its nearest neighbor at a given separation referenced to the likelihood obtained for the same number of galaxies randomly distributed over the same area. Since we are searching for the nearest (i.e. the first) neighbor to each $L^*$ galaxy, these measures are sensitive only to close pairs, not to the large-scale structure as measured in the two-point correlation function.} 
\label{random}
\end{figure}

Comparison of the observed pair distribution with that of a random sample is needed to understand whether the pair distribution could be simply due to chance line-of-sight coincidences. To evaluate the statistical significance of the measured distribution of nearest-neighbor galaxies, we generated samples of galaxies, randomly distributed over the field (excluding masked areas) but with redshift distribution {\it identical} to that of the observed galaxies. For ten repetitions of this Monte-Carlo simulation, we then searched for nearest neighbors at $\Delta z$ within $\pm0.05$ out to a projected physical distance of 150\ts kpc, just as with the real galaxy sample. The number of galaxies in the Monte-Carlo simulations was adjusted to compensate for the variable volumes of the different redshift slices. For the Monte Carlo simulation, the spatial position of each real  
galaxy in the field was randomized but the redshifts were retained; thus any redshift dependent effects were automatically maintained the same for the real galaxy sample and for the simulation galaxies. The ratio of nearest-neighbor distributions for the observed galaxies and the Monte-Carlo simulations is shown in Figure \ref{random} for each redshift range. A completely random set of observed pairs would have a value of unity at all separations in Figure \ref{random}. Instead, the distribution shows a clear excess of neighboring galaxies at small projected separations (and a deficiency at the larger separations). At close separations, the observed/random ratio is above unity and falls off to approximately match the random sample at separations larger than $40-50${\ts}kpc. 

It is important to recognize that since we are searching for the nearest (i.e. the first) neighbor to each $L^*$ galaxy, our measurements are sensitive only to close pairs and they are not sensitive to the large-scale structure as measured in the two-point correlation function. We also note that the Monte-Carlo simulations were entirely random, i.e. without any superposed large scale clustering. Clearly some increases in the frequency of pairs are associated with the non-uniform large scale distribution or clustering of galaxies. But it would be wrong to artificially remove this component of the pair population if it is indeed affecting the actual pair fraction or merger rate. The large-scale structures probably do not have an important effect on the distribution of {\it nearest neighbors} as measured here, although obviously large structures will affect the large-scale correlation function. 

To minimize the contamination from ``random pairs" (i.e. chance line-of-sight superpositions), one should select only those pairs with apparent separations substantially less than 40{\ts}kpc.  For separations $< 20${\ts}kpc, the expected contamination by random pairs is typically $15$\% in each redshift bin (see Figure \ref{random}).  While we could decrease this percentage further by reducing the maximum separation, the total number of real pairs would then decrease, yielding a statistically less-significant sample. A minimum separation of  5{\ts}kpc was also adopted; at smaller separations, the ACS images cannot reliably discriminate cases where internal galaxy structures appear as multiple sources and hence an erroneous pair. 

\section{RESULTS}

The results of our pair identification are summarized in Table \ref{results}. In total, 1,749 galaxy pairs were found across the entire COSMOS field. In Table \ref{results} we separate the numbers of pairs found in ground-based and ACS catalogs in each redshift bin and the expected random component, based on the Monte-Carlo simulations. The pair fraction as given in Table \ref{results} is the fraction of galaxies with a nearest neighbor at 5 to 20 kpc, corrected for the expected random count. Since the pair fraction is calculated using the total number of galaxies in close pairs, in the case of a compact group, each galaxy is counted once, provided it is within 20 kpc of another galaxy. Thus the pair fraction as given here accurately represents the probability that an individual galaxy has a close companion. In cases where the nearest neighbor to one galaxy is a random projection, all galaxies within 20 kpc would still be picked up because they would each have a different nearest neighbor. This ensures that all pairings within 20 kpc are found and then we correct for chance projections as discussed in \S 3.3.

In order to place our results in perspective, we have determined a low redshift galaxy pair fraction using the same criteria adopted for the COSMOS data. The COSMOS survey cannot be used for this ``local" value since the volume sampled in the three lowest redshift bins (i.e. at $z < 0.4$) becomes too small while at the same time, evolution results in there being fewer pairs at low redshift. We were fortunate to be able to use a published catalog of galaxy pairs produced from the SDSS \citep{all04}. From the SDSS catalong, 45 pairs (out of a total of 705 cataloged pairs) at $z<0.1$ met the selection criteria for the projected separations and luminosity (as were used for the COSMOS pairs). A spectroscopic redshift was available for at least one member of each of these pairs but not always for both members of the close pairs. Due to this low level of incompleteness, \citet{all04} estimate that the expected contamination of the sample is $< 3.4\%$.

Figure \ref{fraction} shows the galaxy pair fraction as a function of redshift for the COSMOS pairs together with the local value determined from the SDSS. The error bars shown represent $1\sigma$ Poisson errors. \citet{scoville_cosmos_lss} discuss the cosmic variance and Poisson noise for the COSMOS survey. It is seen there that the cosmic variance decreases relative to the Poisson noise at low masses and should not be significant on the scale of individual galactic halo masses. A simple least squares fit, weighted by the error bars, to the data in Figure \ref{fraction} yields a slope of $n=3.1\pm0.1$. (If the SDSS point is excluded, the best fit power-law has $n=2.8\pm0.1$)  We discuss this result below and compare our findings with previous work. 

\section{DISCUSSION}

The results shown in Figure \ref{fraction} imply strong evolution in the galaxy pair fraction of $>L_V^\star$ galaxies with redshift. The pair fraction rises by a factor of $\sim$10.6 -- from $\sim$1\% at $z = 0.1$ to $\sim$10\% at $z = 1.1$ -- and if extrapolated to higher redshift, would imply that nearly half of all $>L_V^\star$ galaxies at $z \sim 2$ are in close pairs (!)   

The power-law fit shown in Figure \ref{fraction} provides a reasonable representation of the data, with the majority of the data points lying within $2\sigma$ of the straight line of slope $n = 3.1$. The only really significant deviation is for the data point at $z = 0.35$; this turns out to be the redshift bin corresponding to several major galaxy clusters in the COSMOS field \citep{scoville_cosmos_lss}. Thus this high data point is probably high due to an increased pair fraction associated with the overdensity of galaxies in these extended clusters.  Although we discuss the spatial distribution of the identified pairs below, a more thorough investigation of the pair distribution with respect to local environment needs to be carried out to properly address this question; this will be the subject of a second paper.

It is important to note that had our study been limited to a smaller range of redshift, smaller area, or shallower depth, the value of $n$ might have been significantly different due to cosmic variance and Poisson statistics. For example, our sample, limited to $z < 0.5$, would have found $n \sim 5$ while the intermediate redshifts $z = 0.3 - 0.8$ would have yielded $n \sim 0$. Reduced volumes and more limited redshift ranges can be severely affected by counting statistics and large-scale structure excesses or deficiencies (i.e. cosmic variance). The COSMOS survey provides superb angular resolution and depth with the COSMOS ACS and deep ground-based images with excellent seeing over the very large field.  In addition, the availability of a robust SDSS data point allows us to pin down the local value for the pair fraction.  

\subsection{Morphology and Spectroscopy of Pairs} 

\begin{figure}

\plotone{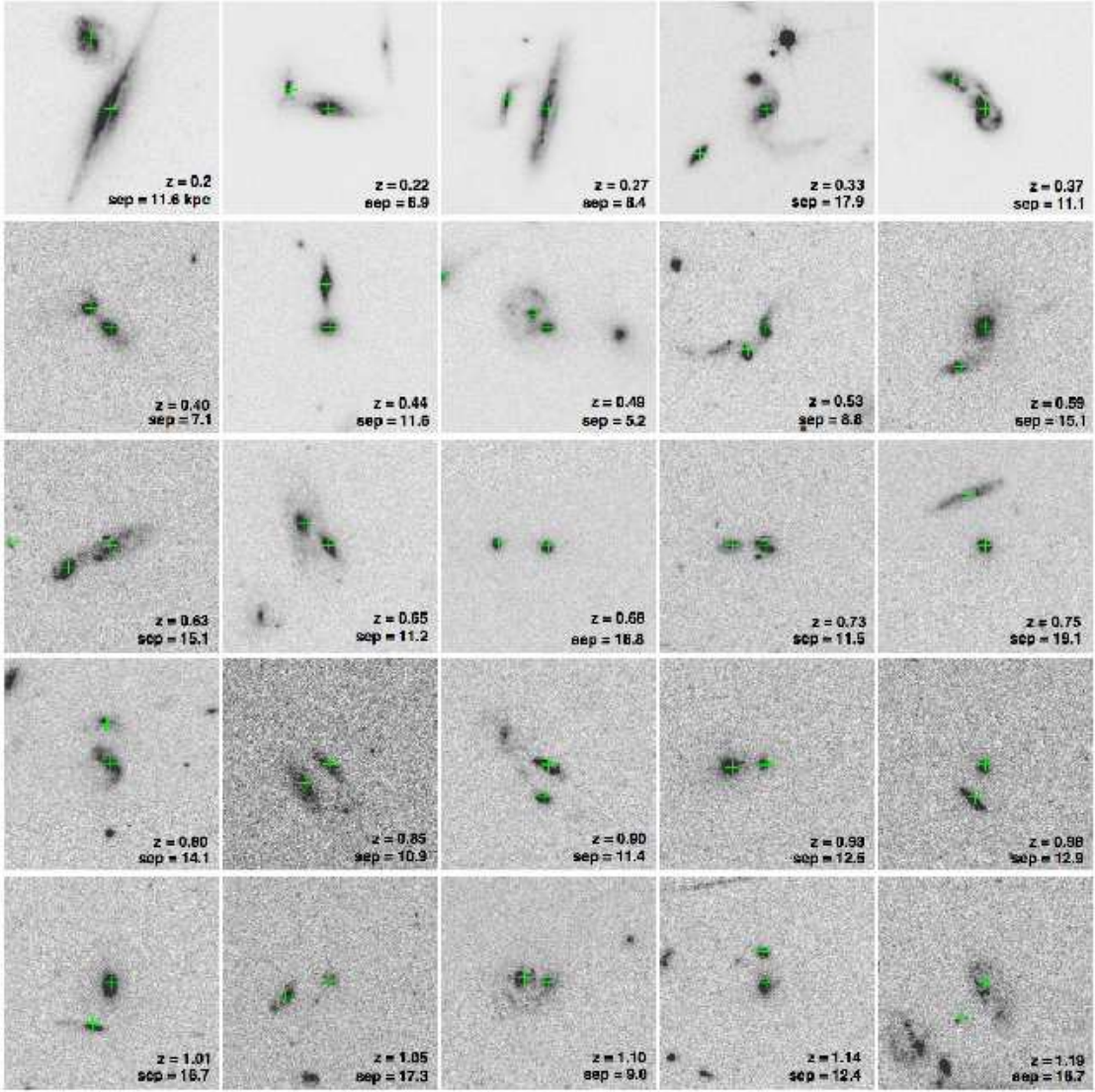}
\vskip 0.4truein
\caption{ACS I-band (F814W) postage stamps ($10\arcsec \times10\arcsec$) of a representative sample of 25 luminous ($\geq L_{\rm V}^\star$) galaxy pairs with projected separations $< 20$ kpc in the COSMOS field ordered by their mean photometric redshift. The redshift of the pair and projected separations in kpc are shown in the lower right corner of each panel. The ACS source positions for 
each member of the pair are marked by green crosses.}
\label{postagestamps}
\end{figure}

To validate the sample of galaxy pairs identified here, we visually inspected the ACS images of all pairs. Figure \ref{postagestamps} shows the ACS images for a sample of 25 galaxy pairs at a variety of separations ordered by their redshifts (but selected at random). The projected separation along with the mean photometric redshift for the pair is shown in the lower right corner of each panel.  Similar plots were made in order to visually inspect all of the 1,749 galaxy pairs in the COSMOS field.   These plots were initially made for the purpose of screening out spurious detections (due to residual cosmic rays, etc.); however, the depth and resolution of the ACS images also makes them extremely useful for discussing the morphology of the galaxies. A substantial fraction of these pairs are obviously interacting, showing clear signs of disturbed morphologies (tidal bridges, tails, etc.). Their morphologies are indicative of early/mid-stage major mergers, involving relatively equal mass pairs, as judged by comparison with the morphologies of strongly interacting pairs at low redshifts.

We are also in the process of obtaining spectra for the complete sample of pairs. Optical spectra for a small subsample of pairs have recently been obtained using DEIMOS on KeckII. A preliminary analysis of these data indicates that $>$85\% of our photometrically identified pairs are indeed galaxies with $\Delta V < 600${\ts}km{\ts}s$^{-1}$, which is in agreement with our expectation based on the expected numbers of observed vs. random pairs shown in Figure \ref{random}. (The complete catalog of objects and their images and spectra will be made available electronically in a future paper.)

\subsection{Spatial Distribution of Pairs} 

\begin{figure}
\plotone{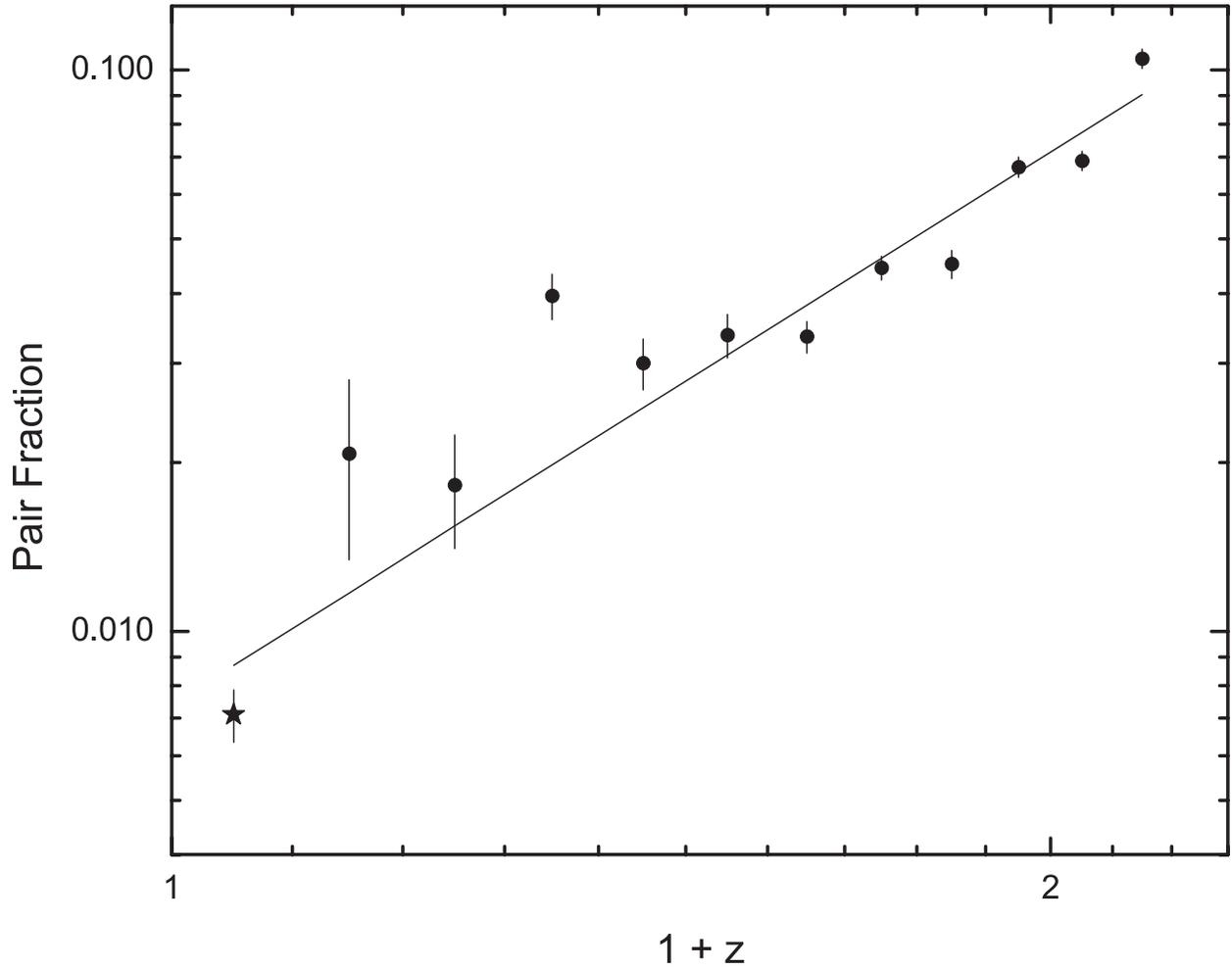}
\vskip 0.4truein
\caption{Fraction of $\geq L_V^\star$ galaxies in close pairs ($5-20$ \ts kpc) versus ($1+z$) for the COSMOS field.  The star marks the local ($z = 0-0.1$) data point determined using data from the SDSS (see text). Vertical bars represent $1\sigma$ error. The straight line least squares fit to the data has a slope of $n=3.1\pm0.1$. }
\label{fraction}
\end{figure}

\begin{figure}
\epsscale{0.90}
\plotone{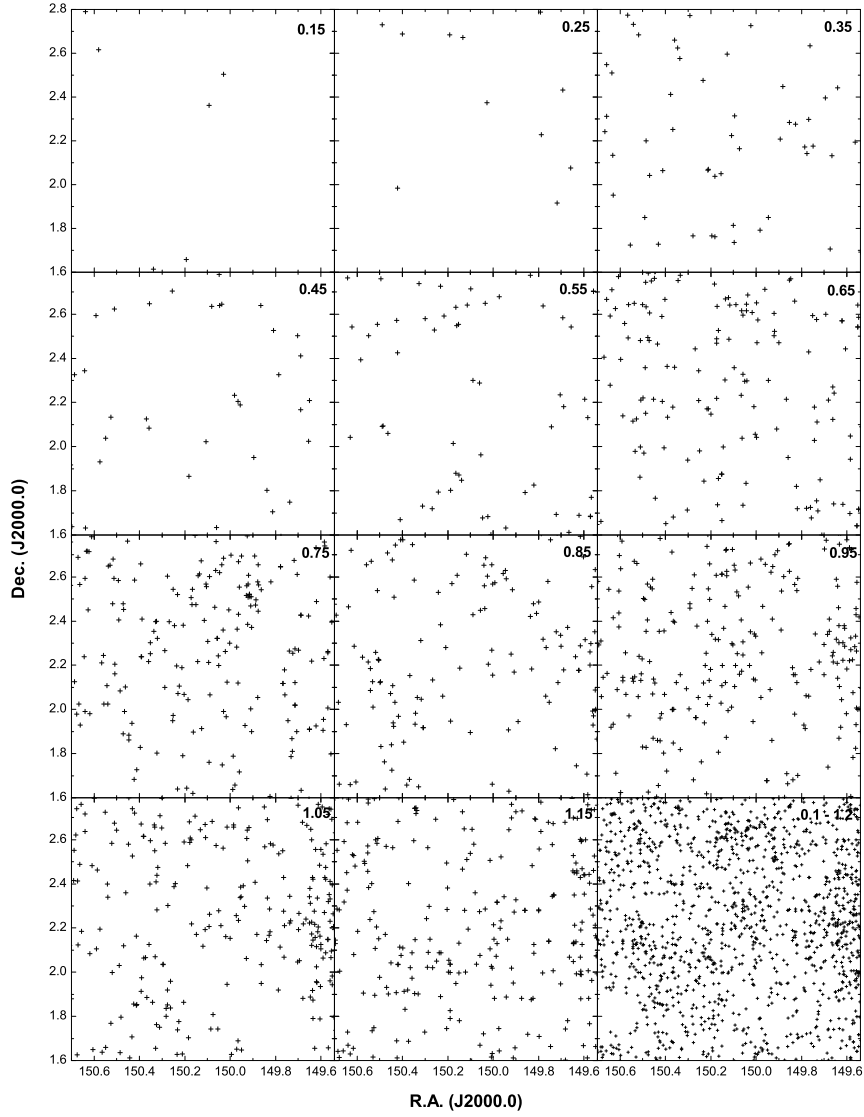}
\vskip 0.2truein
\caption{Distribution of galaxy pairs across the entire 2 \sq$^\circ$\  COSMOS field in $\Delta z=0.1$ redshift bins.  The + symbols represents the centroid position of each identified pair. }
\label{distribution}
\end{figure}

Figure \ref{distribution} shows the distribution of all galaxy pairs found in the COSMOS field in each of the $\Delta z = 0.1$ redshift bins. The distributions of pairs are clearly not random across the field. The most massive large-scale structure seen in the COSMOS field at $z\sim 0.73$ lies in the $0.7 \leq z < 0.8$ redshift bin and a higher density of pairs can be seen in the northwest of the field in this panel in Figure \ref{distribution}. Similarly, there is a significant excess of pairs in the $z=1.05$ slice on the west side of the field. At the higher redshifts ($z>0.5$), the COSMOS field contains very large volumes in each redshift slice so that the overall contribution of a single structure or cluster is diluted and contributes $\leq 10\%$ to the total counts; at lower redshifts, an increase in the pair fraction within large clusters (or alternatively a decrease in the pair fraction in large voids) can cause a larger deviation in Figure \ref{fraction} simply due to the smaller comoving volume of the field. This could partially account for the higher count seen in the $z\sim0.35$ bin (Figure \ref{fraction}). However, the overall number of pairs in the low z slices is also low (due to both the small comoving volume and reduced pair fraction at low redshift), hence, the deviations can also be largely due to small number statistics. In order to assess the significance of cosmic variance across the field, we attempted to divide the field into separate quadrants which were then individually analyzed. Unfortunately, the results in the separate quadrants were even more strongly affected by small number statistics and the quadrant analysis is even less meaningful in assessing cosmic variance. For now, we simply note that one may need to sample volumes at least as as large as $10^6$\ts Mpc$^3$ in order to average out the variations due to LSS, something that COSMOS was designed to do at $z > 0.5$, as the SDSS does at $z < 0.1$.  Nearly the same slope ($n \sim 3.1$) would have been found for the data in Figure \ref{fraction} had only the SDSS data point and the COSMOS data at $z > 0.5$ been used in the fit.

\subsection{Comparison with Previous Studies}

To facilitate comparison of our results for evolution of the galaxy pair fraction ($\propto(1+z)^{3.1\pm0.1}$) with prior studies, we compile in Table \ref{summary} a summary of earlier studies of pair fraction and merger rate evolution. This table also includes several recent studies which used morphological criteria presumed to be related to interactions and mergers as a basis for directly sampling systems that have interacted and subsequently merged \citep[e.g.,][]{lef00,con03,lav04,lot06}. A critique of both methods (pair fraction and disturbed morphologies) is provided in \citet{lav04}. The results of both are usually parameterized as a power-law in $(1+z)$.  Although it is reasonable to believe that the two power laws would be related, the indices are probably not the same since not all pairs will merge and their merging timescales are different. The fraction of pairs which do eventually merge must depend on their separations and the distribution of separations is likely to vary with redshift due to merging. Several studies have argued that power-law index $m$ for the merger rate ranges between $\pm 1$ from the index $n$ used to parameterize the pair fraction \citep[e.g.,][]{lav04}. For samples like ours, involving pairs of {\it luminous} (i.e. $\geq L_V^\star$) and presumably massive galaxies at {\it close} separations ($\leq 20$kpc), we expect that $(m-n)\sim0$ since most of these systems should eventually merge.

Most of the investigations listed in Tables \ref{summary} used spectroscopic redshifts (at least for objects at $z \la 1$) so the samples are relatively small -- typically $\sim 2-8$\% the size of our COSMOS sample. These samples may also be incomplete at high redshift where spectroscopic surveys are unable to resolve very close pairs.  While some of these studies suffer serious incompleteness due to their apparent magnitude limit, several of the most recent, using deep surveys, have sufficient sensitivity to adopt an absolute magnitude selection limit similar to what we have done here. The derived values of $m$ and $n$ span an enormous range -- from $0-6$. While ``$m$ versus $n$" can account for some (i.e. $\pm 1$) of this variation, it is more likely that different redshift ranges, different ways of correcting for non-physical (line of sight) galaxy pairs, cosmic variance, and small sample sizes are the major contributors to the large discrepancies in power-law indices. The straight average of all measurements in Table \ref{summary} (excluding the current work) is $\sim 2.1\pm0.8$, with the mean for methods using pair-fraction and for methods using morphology being nearly equal. One notable difference concerns those studies that have adopted ``passive luminosity evolution" (PLE) models to ``correct" the observable luminosity measurements to mass measurements.  These more recent studies (flagged in Table \ref{summary}) give systematically smaller values, with an average of $1.0\pm0.3$.  

\subsection{Galaxy Masses and Luminosities:  Passive Luminosity Evolution (PLE)?}

Many recent studies of the merger rate have assumed passive (or pure) luminosity evolution (PLE) in an attempt to select similar galaxies at all redshifts. PLE is typically parameterized by magnitudes $M(z)=M(z=0)-Qz$ where the empirical assumption has been that $Q \simeq 1$ \citep{lin04,pat02}. We tested such a model, selecting galaxies using $M_V=-19.8$ in our highest redshift bin ($1.1<z\leq1.2$), but including fainter galaxies at the lower redshifts, assuming $Q=1$ in the equation above. The rest of the selection criteria remained the same, resulting in a sample of 117,015 galaxies. (The sightly increased number counts arise in the lowest redshift bins.) Using this PLE sample of galaxies and counting the pairs again, we obtained a fit with $n = 2.2\pm0.1$. 

However, we do not believe that PLE is justified empirically. To test for evolutionary variation in the ratio $M/L_V$ as a function of redshift, we made use of the galaxy masses derived from the COSMOS photometry. To estimate the mass of each galaxy in the COSMOS photometric catalog, \citet{mobasher_cosmos_photz} used a relation between $M/L_V$ and rest-frame $(B-V)_0$ colors given by \citet{bel05}. These masses are shown for the pair sample galaxies in Table \ref{Masses}. They are consistent with relatively little or no variation in mean mass versus redshift for our pair sample. The median spectral type of the galaxy pairs indicates that they are late-type spirals for which passive evolution is inappropriate since these late type galaxies have continuous, on-going star formation. The apparent difference in slope is more likely due to an increased merger fraction at faint magnitudes than passive evolution for the massive galaxy pairs. In summary, {\it we find no basis for adopting PLE for the COSMOS sample of galaxy pairs and instead, retain the fixed $L_V^\star$ sample as most relevant 
for analysis of the evolution of galaxy merging}. 

\section{SUMMARY}
COSMOS HST and ground-based images covering the full 2{\ts}\sq$^\circ$ COSMOS field have been used to identify an unprecedented sample of 1,749  luminous ($\geq L_V^\star$) galaxy pairs at redshift $z \leq 1.2$. These data, supplemented with a complete sample of 45 ``local" galaxy pairs selected from the SDSS, have been used to determine the pair fraction of luminous galaxy pairs as a function of redshift over the range $0 < z \leq 1.2$.  In addition, galaxy masses obtained from an analysis of the extensive multi-wavelenghth (UV-opt-NIR) COSMOS data, have been used to investigate possible evolution in $M/L$ versus redshift. From our analysis, the following conclusions are drawn:  

\begin{enumerate}
	\item Evolution of the galaxy pair fraction of close pairs (with projected separation of $5-20$ \ts kpc) can be fit as  $(1+z)^n$, where we find $n = 3.1 \pm 0.1$.
	
	\item The HST-ACS images suggest that a large majority of  our luminous galaxy pairs are strongly interacting/merging disks galaxies.
	
	\item The mean $M/L$ ratio of the galaxy pairs in our sample shows no evidence of evolution with redshift. This argues against the use of {\it passive evolution} models in conjunction with the sample selection. (In any case, it is hard to justify using passive evolution models for star forming disk galaxies which are typically the focus of merger galaxy studies).
	
\end{enumerate}

The high angular resolution and large area coverage provided by the COSMOS HST-ACS imaging were crucial in allowing us to cleanly separate close pairs (separations $<20${\ts}kpc) out to $z=1.2$ and providing an enormous sample of 1749 galaxy pairs -- $10-20 \times$ larger that that used in previous studies. The ability to focus the study on close pairs minimized the number of chance line-of-sight pairs and increased our ability to identify strongly interacting/merging systems. The very large sample size was critical for reducing cosmic variance and Poisson errors; these variances have been a major problem for earlier studies.

The resolution of the ACS images would, in principle, allow us to extend the investigation to $z \sim 2$ once deeper near-infrared  ($Y,J,H,K_{\rm s}$) and {\it  Spitzer-}IRAC (3.6, 4.5, 5.6, 8.0 $\mu$m) data are available for the photometric redshift analysis.  The results presented here show that $\sim 10\%$ of all luminous ($\geq L_V^\star$) galaxies are in close pairs already by $z=1.2 $; extrapolating the $(1+z)^{3.1}$ power-law to higher redshift suggests that $\sim$50\% of all luminous galaxies are in close pairs by $z=2$.  Thus major interactions and mergers must represent an important and very likely dominant process of galaxy evolution in the early universe.

\acknowledgments
The HST COSMOS Treasury program was supported through NASA grant HST-GO-09822.  We gratefully acknowledge the contributions of the entire COSMOS collaboration consisting of more than 70 scientists. We also wish to thank the referee for helpful comments that have greatly improved the presentation of the data in our paper. JSK would like to thank Lisa Kewley for many helpful discussions concerning this project as well as Sahar Allam for providing the catalog of SDSS galaxies that her pair sample was drawn from. NZS would like to thank the Institute for Astronomy, University of Hawaii for support during a sabbatical stay where this project began. DBS and NZS also acknowledge the hospitality of the Aspen Center for Physics.

\nocite{car94,neu95,pat97,aus07,koe07,res00,lea_cosmos_catalog}

\bibliographystyle{apj}
%\bibliography{mergers,cosmos}

\clearpage
\begin{deluxetable}{lccccccccccc}
  \tablewidth{0pt}
  \tabletypesize{\scriptsize}
  \tablecaption{Number of COSMOS Galaxies ($\geq L_V^\star$) with a Companion at Projected Separation $5-20${\ts}kpc \label{results}}
  \tablehead{  \colhead{$z$\tablenotemark{a} } & \colhead{0.15} & \colhead{0.25} & \colhead{0.35}&
  \colhead{0.45} & \colhead{0.55} & \colhead{0.65} & \colhead{0.75} & \colhead{0.85} &
  \colhead{0.95} & \colhead{1.05} & \colhead{1.15}}
    \startdata
     ACS only &       0 &  0 &   0 &   0 &   0 &   5 &  9 &   14 &  44 &  85 &  59 \\       
     Ground Based &  10 & 27 & 148 & 117 & 163 & 381 & 597 & 398 & 757 & 772 & 837 \\      
     Total &         10 & 27 & 148 & 117 & 163 & 386 & 606 & 412 & 801 & 857 & 896 \\
     Total - Random & 8 & 19 & 117 &  93 & 124 & 244 & 424 & 299 & 584 & 612 & 642 \\
     Galaxy Sample & 386 & 1043 & 2956 & 3097 & 3680 & 7320 & 9575 & 6719 & 8888 & 9276 & 6281 \\
     Pair Fraction & 0.021 & 0.018 & 0.040 & 0.030 & 0.034 & 0.033 & 0.044 & 0.045 & 0.066 & 0.066 & 0.102 \\
     Comoving Volume sampled ($10^6$\ts Mpc$^3$)& 0.09 & 0.23 & 0.41 & 0.61 & 0.81 & 1.01 & 1.21 & 1.38 & 1.54 & 1.69 & 1.82 \\
     \enddata
     \tablenotetext{a}{Centroid of redshift bin of width 0.1}
\end{deluxetable}

\begin{deluxetable}{lcccccc}
  \noindent
 \tablewidth{0pt}
  \setlength{\tabcolsep}{0.05in}
  \tabletypesize{\scriptsize}
  \tablecaption{Summary and Comparison with Previous Work \label{summary}}
  \tablehead{\colhead{Reference} & \colhead{mag\tablenotemark{a}} & 
  \colhead{$z$\tablenotemark{b}} & \colhead{$z$ Range\tablenotemark{c}} & \colhead{Sample Size\tablenotemark{d}} &
  \colhead{\#P or M\tablenotemark{e}} & \colhead{$m$ or $n$ \tablenotemark{f}}}
    \startdata
    \sidehead{\it Galaxy Pair Fraction Studies}

     Zepf \& Koo (1989)        & $B\leq22$     & A     & $<z>=0.25$    & $\sim1000$ & 20  & $4.0\pm2.5$ \\
     Carlberg et al. (1994)    & $V\leq22.5$   & S     & $<z>=0.42$    & 1062       & 14  & $2.3\pm1.0$ \\
     Burkey et al. (1994)      & $I$=18.5-23   & A,S   & $0.4-0.7$     & 146        & 50  & $3.5\pm0.5$ \\
     Woods et al. (1995)       & $I\leq24$     & N     &    \nodata           & $\sim1000$ & 23  & 0 \\
     Neuschaefer et al. (1995) & $I\leq22$     & \nodata & $z_{med}=0.5$ & $\sim4500$ &  \nodata & 0 \\
     Yee \& Ellingson (1995)   & $r\leq21.5$   & S     & $<z>=0.38$    & 107        & 25  & $4.0\pm1.5$ \\
     Neuschaefer et al. (1997) & $I$=18-23     & S     & $\la 1-2$     & $\sim22400$& 90  & $1.2\pm0.4$ \\
     Patton et al. (1997)      & $r\leq22$     & S     & $<z>=0.33$    & 545        & 73  & $2.8\pm0.9$ \\
     Wu \& Keel (1998)         & $I\leq24$     & A     & $\sim 2$      &     \nodata       & 10  & $0-2$  \\
     Carlberg et al. (2000)    & $M_R\leq19.8$ & S     & $0.1-1.1$     & $\sim$3300 & 109 & $0.1\pm0.5$ \tablenotemark{g}  \\
     Le Fevre et al. (2000)    & $\delta m\leq1.5$ & S & $\leq 1$      & 285        & 26 & $2.7\pm0.6$ \\
     Patton et al. (2002)      & $R_c\leq21.5$ & S     & $0.12-0.55$   & 4184       & 88   & $2.3\pm0.7$\tablenotemark{g}  \\
     Lin et al. (2004)         & $M_B$=-(21-19)& S     & $0.45-1.2$    & 5000       &  \nodata  & $1.60\pm0.29$ \\ 
     Lin et al. (2004)         &               &       &    &      &  \nodata  & $0.51\pm0.28$ \tablenotemark{g} \\
     \\
     {\bf This work \tablenotemark{h}} & {\bf $\bf >L_{V}^{\star}$} & \bf Ph     & $\bf 0-1.2$    & \bf 106118     & \bf 1749& $\bf 3.1\pm0.1$ \\
\\
    \sidehead{\it Morphologically Selected Merger Studies}
     Le Fevre et al. (2000)    & $\delta m\leq1.5$ & S & $\leq 1$      & 285        & 49 & $3.4\pm0.6$ \\
     Reshetnikov (2000)        & \nodata       & Ph    & $0.5-1.0$     & \nodata    & 14   & $3.6^{+1.2}_{-0.9}$ \\
     Conselice et al. (2003)   & $M_B\leq-18$  & Ph,S  & $0-3$         & 1212       & 43   & $4-6$   \\
     Lavery et al. (2004)      & \nodata       & Ph,S  & $0.1-1.0$     & \nodata & 25 & $5.2\pm0.7$ \\
     Lotz et al. (2006)        & $>0.4L_{B}^{\star}$& S     & $0.2-1.2$    & 2368       & 157& $1.12\pm0.60$\tablenotemark{g} 
     
    \enddata 
    \tablenotetext{a}{Magnitude selection limits as stated in the reference. $\delta m$ refers to difference in magnitude and $R_c$ is a Kron-Cousins R magnitude.}
    \tablenotetext{b}{Method used for redshift measurement:  S = spectroscopic, Ph = photometric,  A = assumed,  N = none}
    \tablenotetext{c}{Redshift range as stated in the reference. In some cases, only the mean, median, or upper limit of the redshift range was given and then compared to a local value.}
    \tablenotetext{d}{Galaxy sample size, as stated in the reference.  In the case of \citet{lav04}, 162 WFPC2 fields were used. }
    \tablenotetext{e}{Total number of identified Pairs (P) or morphologically selected Mergers (M) in the sample.}
    \tablenotetext{f}{Exponent: merger rate ($m$) or pair fraction ($n$), assuming evolution $\propto (1+z)^{m,n}$. Error bars are assumed to be $1\sigma$.}
    \tablenotetext{g}{Assuming passive luminosity evolution.}
    \tablenotetext{h}{Values for total number of galaxies and pairs includes only the COSMOS data. Including the SDSS data used for the $z=0-0.1$ data point would only increase these values}
\end{deluxetable}

\begin{deluxetable}{lccccccccccc}
  \tablewidth{0pt}
  \tabletypesize{\scriptsize}
  \tablecaption{Log of Stellar Masses of Galaxy Pairs \label{Masses}}
  \tablehead{  \colhead{$z$\tablenotemark{a} } & \colhead{0.15} & \colhead{0.25} & \colhead{0.35} &
  \colhead{0.45} & \colhead{0.55} & \colhead{0.65} & \colhead{0.75} & \colhead{0.85} &
  \colhead{0.95} & \colhead{1.05} & \colhead{1.15}}
    \startdata
		Mean          & 10.03 & 10.04 & 10.07 & 10.11 & 10.15 & 10.14 & 10.11 & 10.15 & 10.13 & 10.13 & 10.09 \\ 
		Median        &  9.83 &  9.98 & 10.02 & 10.11 & 10.17 & 10.11 & 10.09 & 10.13 & 10.12 & 10.12 & 10.07 \\
		Stddev        &  0.48 &  0.43 &  0.45 &  0.42 &  0.47 &  0.45 &  0.46 &  0.44 &  0.44 &  0.46 &  0.45 \\
   \enddata
  \tablenotetext{a}{Centroid of redshift bin of width 0.1}
\end{deluxetable}
\clearpage

\end{document}